\documentclass[twocolumn,preprintnumbers,amsmath,aps]{revtex4}
\usepackage{graphicx}
\usepackage{dcolumn}
\usepackage{bm}
\usepackage{subfigure}
\usepackage[usenames]{color}
\begin{document}
\newcommand{\kvec}{\mbox{{\scriptsize {\bf k}}}}
\def\eq#1{(\ref{#1})}
\def\fig#1{figure\hspace{1mm}\ref{#1}}
\def\tab#1{table\hspace{1mm}\ref{#1}}
\title{
Study of the superconducting phase in silicene under biaxial tensile strain}
\author{A.P. Durajski$^{1}$}\email{adurajski@wip.pcz.pl}
\author{D. Szcz{\c{e}}{\'s}niak$^{2}$}
\author{R. Szcz{\c{e}}{\'s}niak$^{1}$}
\affiliation{1. Department of Solid State Physics, Institute of Physics, Cz{\c{e}}stochowa University of Technology, Ave. Armii Krajowej 19, 42-200 Cz{\c{e}}stochowa, Poland}
\affiliation{2. Department of Theoretical Physics, Institute of Physics, Jan D{\l}ugosz University, Ave. Armii Krajowej 13/15, 42-200 Cz{\c{e}}stochowa, Poland}
\date{\today} 
\begin{abstract}
The electron-doped silicene under the influence of the biaxial tensile strain is predicted to be the phonon-mediated superconductor. By using the Eliashberg formalism, we investigate the thermodynamic properties of the superconducting silicene in the case when the tension is $5\%$ and the electron doping equals $3.5\times10^{14}~{\rm cm^{-2}}$. Under such conditions, silicene monolayer is expected to exhibit the highest superconducting transition temperature ($T_C$). In particular, based on the electron-phonon spectral function and assuming wide range of the Coulomb pseudopotential values ($\mu^{\star}\in\left\langle0.1,0.3\right\rangle$) it is stated that the superconducting transition temperature decreases from $18.7$ K to $11.6$ K. Similar behavior is observed in the case of the zeroth temperature superconducting energy gap at the Fermi level: $2\Delta(0)\in\left\langle6.68, 3.88\right\rangle$ meV. Other thermodynamic parameters differ from the predictions of the Bardeen-Cooper-Schrieffer theory. In particular, the ratio of the energy gap to the critical temperature changes in the range from $4.14$ to $3.87$. The ratio of the specific heat jump to the specific heat in the normal state takes the values from $2.19$ to $2.05$, and the ratio of the critical temperature and specific heat in the normal state to the thermodynamic critical field increases from $0.143$ to $0.155$. It is also determined that the maximum value of the electron effective mass equals $2.11$ of the electron band mass.
\end{abstract}
\maketitle

\section{Introduction}
Recent research in the field of nanotechnology have led to the synthesis and characterization of various two-dimensional materials \cite{Castro}. The unique geometries of these novel structures are one of the main origins of their extraordinary physical and chemical properties \cite{Castro}, \cite{Xu}. Among different applications, the possibility of using such low-dimensional systems in the domain of nanoscale superconducting devices is of growing interest \cite{Franceschi}, \cite{Delahaye}, \cite{Saira}.

In this respect, the one-atom-thick two-dimensional form of carbon, known as graphene \cite{Novoselov}, attracted exceptional attention in recent years, when comparing to the other carbon allotropes \cite{Hirsch}, \cite{Szczesniak1}, \cite{Yang}. However, various theoretical calculations demonstrate that the phonon-mediated superconductivity does not occur in the intrinsic graphene, due to the weak electron-phonon coupling constant \cite{Forti}, \cite{Johannsen}, \cite{Zhang}. This fact follows the case of graphite, where the induction of the superconducting phase is possible only via the chemical process know as intercalation \cite{Profeta}.

In particular, it was suggested, by using the first-principle calculations, that the conventional superconducting state with notable high critical temperature ($T_C$) can be observed in the hole-doped graphane (a fully hydrogenated graphene) \cite{Sofo}, \cite{Savini} or in the lithium-decorated graphene \cite{Profeta}, \cite{LiC6}, \cite{Kaloni}. Due to these findings this direction of research can be considered as a promising and important one. However, the direct experimental evidences are still lacking.

Another noteworthy material for the superconducting nanoelectronic applications is silicene (the graphene-like monolayer of silicon) \cite{Aufray}. In general, the main advantage of this material is the fact that it combines some of the graphene intriguing properties and can be relatively easy incorporated into the existing silicon-based electronics \cite{Jose}. Moreover, the preliminary results on the synthesis of silicene are already available \cite{Vogt}.

From the point of view of the superconducting properties, it is important that pristine silicene is characterized by the buckled structure, which distinguish it from the graphene and favours stronger electron-phonon coupling in this material \cite{Wan}. 
Some theoretical works, using random-phase-approximation (RPA), have proposed possible singlet $d+id'$ chiral supercondutivity in undoped bilayer silicene \cite{FengLiu} or quantum phase transition to the triplet $f-$wave superconducting phase in doped silicene under a perpendicular external electric field \cite{Li-DaZhang}.
Encouraging are also recent experimental results which suggest that the induction of the superconducting state in supported silicene may be possible \cite{Chen}.

On the other hand, latest theoretical investigations predict that the electron-doping and the influence of the biaxial tensile strain induce superconducting state characterized by the critical temperature much above 10 K \cite{Wan}. In particular, for the electron doping ($n_e$) equals $3.5\times10^{14}~{\rm cm^{-2}}$ and tension of $5\%$, the analytical McMillan \cite{McMillan} formula gives $T_C$ $\sim 17$ K. This outcome is promising and motivates our studies.

In the present paper, we concentrate ourselves on the analysis of the superconducting phase induced in silicene under the conditions mentioned above. In the considered case, the electron-phonon coupling constant exceeds the weak coupling limit ($\lambda>0.5$ \cite{Bauer}). Due to this fact we conduct our calculations within the framework of the Eliashberg formalism \cite{Eliashberg}, which allows us to calculate the thermodynamic properties of the superconducting phase at the quantitative level. Our calculations based on the electron-phonon spectral function $\alpha^2F(\omega)$ obtained in \cite{Wan}, by using the density functional theory within the local-density approximation. The numerical methods adopted in the present work based on the self-consistent iterative procedure presented in \cite{LiC6}, \cite{GaH3}, \cite{Domin}.

\section{The numerical and analytical results}

In order to compute all thermodynamic properties of interest, the Eliashberg equations are solved on the imaginary axis and in the mixed representation (defined simultaneously on the imaginary and real axis). The stability of the solutions in both cases is achieved in the temperature range from $T_0=1$ K to $T_C$, for the assumed $1100$ Matsubara frequencies: $\omega_{m}\equiv\frac{\pi}{\beta}(2m-1)$, where $\beta\equiv 1/k_{B}T$, and $k_{B}$ denotes the Boltzmann constant. 

The Coulomb pseudopotential ($\mu^{\star}$) models the depairing interaction between the electrons and beside the Eliashberg function is the second input parameter in the Eliashberg equations.
In fact there are two well-known experimental methods to determine $\mu^{\star}$. One is based on the isotope effect \cite{McMillan}, second is based on the inversion of tunnelling data \cite{AllenDynes}.
It should be emphasized, that the isotope effect can be used only when a set of isotopes is available and the tunnelling experiments require strong or medium coupling superconductors in order to give sufficient structure in the current-voltage characteristic \cite{Rapp}. 
The physical value of Coulomb pseudopotential is hard to calculate from first-principles, so it is usually treated as a material-dependent adjustable parameter chosen, for examples within the framework of the Eliashberg formalism, to reproduce the experimental value of critical temperature \cite{SzczesniakCoulomb}.
However, due to absence of the experimental results for silicene, a wide range of the Coulomb pseudopotential values is taken into account: $\mu^{\star}\in\left\langle0.1,0.3\right\rangle$.

\begin{figure}[ht]
\includegraphics[width=\columnwidth]{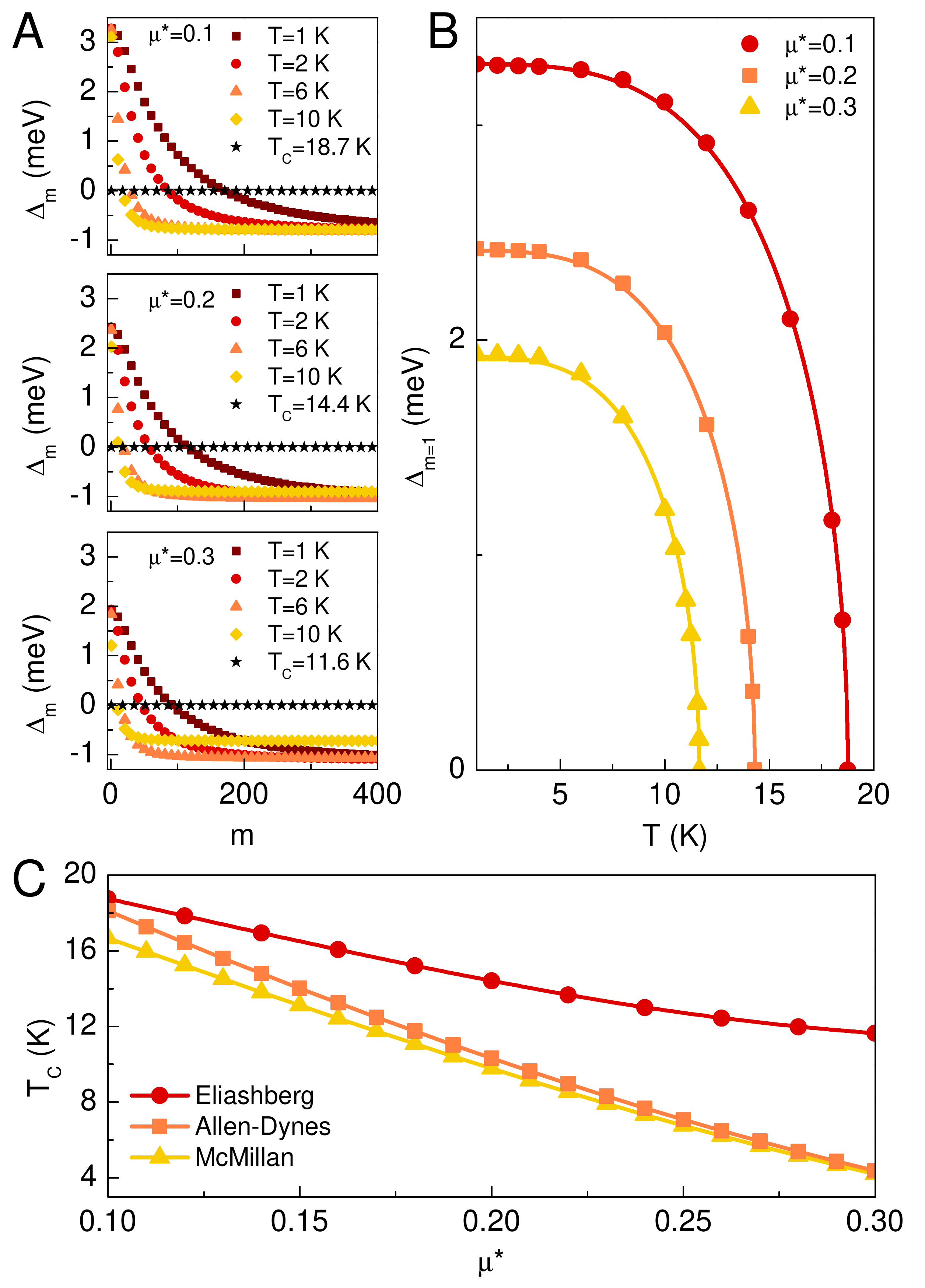}
\caption{(A) The dependence of the order parameter on the number $m$ for the selected temperatures and Coulomb pseudopotential values. (B) The full dependence of the maximum value of the order
parameter on the temperature for selected $\mu^{\star}$. (C) The critical temperature as a function of the Coulomb pseudopotential. The circles correspond to the exact numerical solutions of the Eliashberg equations. The triangles and squares represent the results obtained using the Allen-Dynes and McMillan formula, respectively.}
\label{fig1}
\end{figure}

In \fig{fig1} (A), the superconducting order parameter on the imaginary axis for selected values of the temperature and the Coulomb pseudopotential is presented. The maximum value of the order parameter ($\Delta_{m=1}$) decreases with the growth of $T$ and $\mu^{\star}$. This fact can be clearly observed in \fig{fig1} (B) where $\Delta_{m=1}\left(T\right)$ function is shown. On the basis of these results, we note that the $\Delta_{m=1}$ values can be well parameterized by the following formula:
\begin{equation}
\label{r1}
\Delta_{m=1}\left(T,\mu^{\star}\right)=\Delta_{m=1}\left(\mu^{\star}\right)\sqrt{1-\left(\frac{T}{T_{C}}\right)^{\alpha}},
\end{equation}
where: $\Delta_{m=1}\left(\mu^{\star}\right)=18.30\left(\mu^{\star}\right)^{2}-14.08\mu^{\star}+4.51$ and the fitting parameter $\alpha$ is equal to 3.4. 

The superconducting transition temperature is defined as the temperature at which the order parameter vanishes: $\Delta_{m=1}\left(T_{C},\mu^{\star}\right)=0$. In the case of silicene it is stated that $T_{C}$ is high in the entire range of the Coulomb pseudopotential values. In particular, $T_{C}\in\left<18.7, 11.6\right>$ K for $\mu^{\star}\in\left<0.1, 0.3\right>$. 
It should be noted that the maximum value of the critical temperature for $\mu^{\star}=0.1$ determined by us is significantly higher then the value predicted in paper \cite{Wan}, where $[T_C]^{\rm max}_{\mu^{\star}=0.1}=16.40$ K. This situation is caused by the fact that in paper \cite{Wan} the superconducting transition temperature was estimated by using McMillan formula \cite{McMillan}:

\begin{equation}
\label{r2}
k_{B}T_{C}=\frac{\omega_{\rm ln}}{1.2}\exp\left[\frac{-1.04\left(1+\lambda\right)}{\lambda-\mu^{\star}\left(1+0.62\lambda\right)}\right],
\end{equation}
where $\omega_{\rm ln}$ is the logarithmic averaged phonon frequency and equals $18.52$ meV for the tension of $5\%$.

A full dependence of $T_{C}$ on $\mu^{\star}$ is plotted in \fig{fig1} (C). Presented results are obtained using the Eliashberg formalism, Allen-Dynes expression \cite{AllenDynes} and the McMillan formula \cite{McMillan}. It can be observed that the analytical results underestimate the critical temperature, especially for the high values of the Coulomb pseudopotential. Moreover, the Allen-Dynes expression much better predicts $T_C$ than the McMillan formula.

\begin{figure}[h]
\includegraphics[width=\columnwidth]{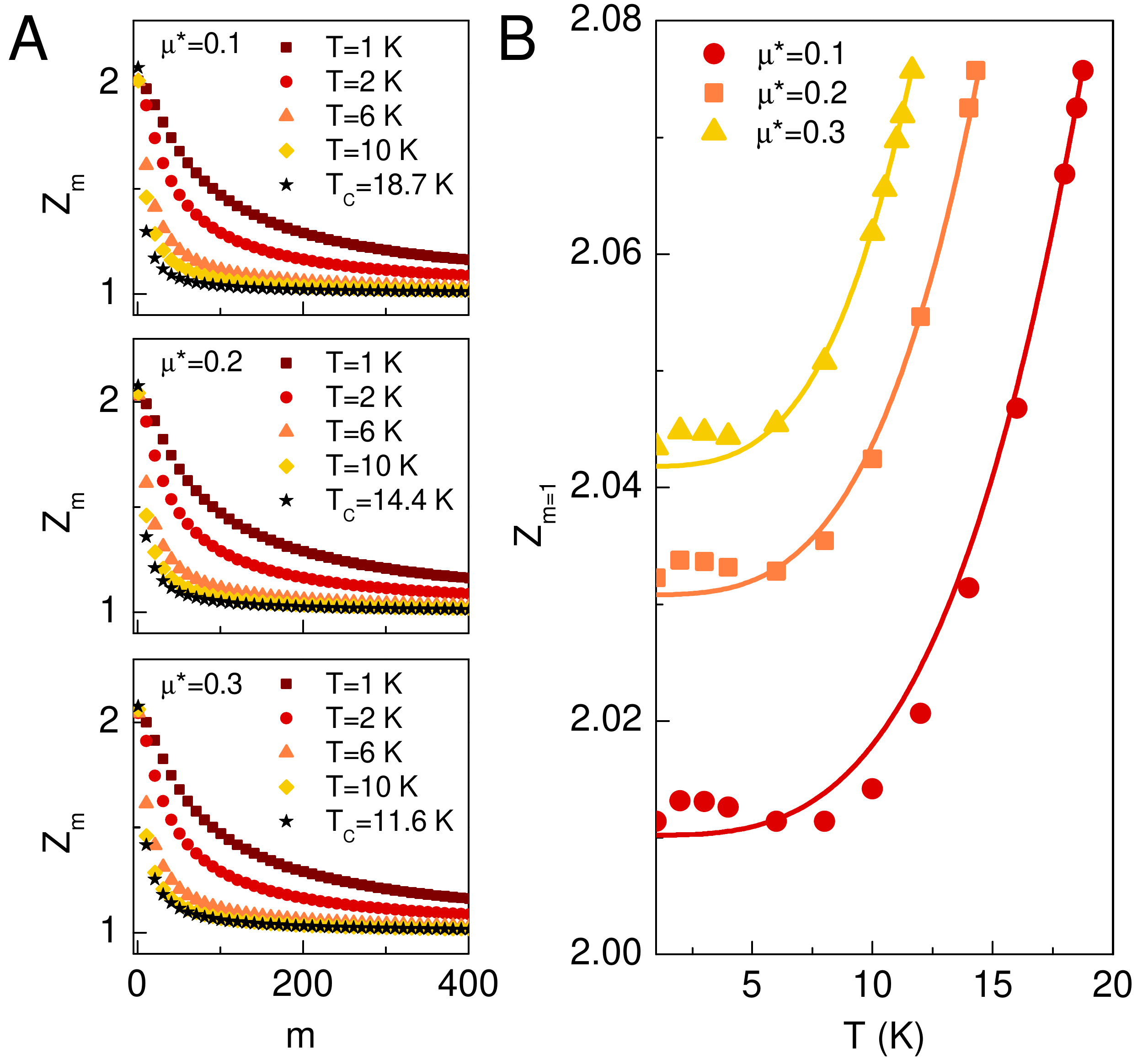}
\caption{(left panel) The form of the wave function renormalization factor on the imaginary axis for selected values of temperature and Coulomb pseudopotential, and (right panel) the dependence of the maximum value of the wave function renormalization factor on the temperature.}
\label{fig2}
\end{figure}

In \fig{fig2}, we present the results for the wave function renormalization factor. The identical values of temperature and Coulomb pseudopotential as for the order parameter are chosen. It is found that the function $Z_{m=1}(T)$ increases together with the temperature and the Coulomb pseudopotential value.

The values of the function $Z_{m=1 }\left(T,\mu^{\star}\right)$ can be estimated by the formula: 
\begin{eqnarray}
\label{r3}
Z_{m=1}\left(T,\mu^{\star}\right)&=&Z_{m=1}\left(\mu^{\star}\right)\\ \nonumber
&+&\left[Z_{m=1}\left(T_{C}\right)-Z_{m=1}\left(\mu^{\star}\right)\right]\left(\frac{T}{T_{C}}\right)^{\alpha},
\end{eqnarray}
where: $Z_{m=1}\left(\mu^{\star}\right)=-0.48\left(\mu^{\star}\right)^{2}+0.35\mu^{\star}+1.98$ and $Z_{m=1}\left(T_{C}\right)=1+\lambda$. Note that for $T=T_C$, the maximum value of the wave function renormalization factor is independent of $\mu^{\star}$.

The temperature dependence of $\Delta_{m=1}$ is reflected in measurable thermodynamic parameters as the specific heat $C^{S}(T)$ of the superconducting state or the thermodynamic critical field $H_C(T)$. In order to investigate these properties from the solution of Eliashberg equations on imaginary axis, we have evaluated numerically the free energy difference between the superconducting and the normal state ($\Delta F=F^{S}-F^{N}$). This is given by the formula \cite{BardeenStephen}:
\begin{eqnarray}
\label{r4}
\frac{\Delta F}{\rho\left(0\right)}&=&-\frac{2\pi}{\beta}\sum_{n=1}^{M}
\left(\sqrt{\omega^{2}_{n}+\Delta^{2}_{n}}- \left|\omega_{n}\right|\right)\\\nonumber
&\times&\left(Z^{S}_{n}-Z^{N}_{n}\frac{\left|\omega_{n}\right|}
{\sqrt{\omega^{2}_{n}+\Delta^{2}_{n}}}\right),
\end{eqnarray}
where $\rho\left(0\right)$ is the value of the electron density of states at the Fermi energy level. Symbols $Z^{S}_{n}$ and $Z^{N}_{n}$ denote the wave function renormalization factors for the superconducting ($\Delta_{m=1}\neq 0$) and for the normal state ($\Delta_{m=1}= 0$), respectively. 

The thermodynamic critical field and deviation function of the thermodynamic critical field are calculated from the free energy difference:
%
\begin{equation}
\label{r5}
\frac{H_{C}}{\sqrt{\rho\left(0\right)}}=\sqrt{-8\pi\left[\Delta F/\rho\left(0\right)\right],}
\end{equation}
%
\begin{equation}
D = {H_c\left(T\right)\over{H_c\left(0\right)}} -
\Bigl[1-\Bigl({T\over{T_c}}\Bigr)^2\Bigr]\,.
\label{rD}
\end{equation}

The specific heat difference $\Delta C=C^{S}-C^{N}$ is related to $\Delta F$ through:
\begin{equation}
\label{r6}
\frac{\Delta C\left(T\right)}{k_{B}\rho\left(0\right)}=-\frac{1}{\beta}\frac{d^{2}\left[\Delta F/\rho\left(0\right)\right]}{d\left(k_{B}T\right)^{2}},
\end{equation}
while the specific heat in the normal state is defined as: $C^{N}=\gamma/{\beta}$, where $\gamma$ is the Sommerfeld constant: $\gamma\equiv ({2}/{3})\pi^{2}\left(1+\lambda\right)k_{B}\rho\left(0\right)$. 

\begin{figure}[hb]
\includegraphics[width=\columnwidth]{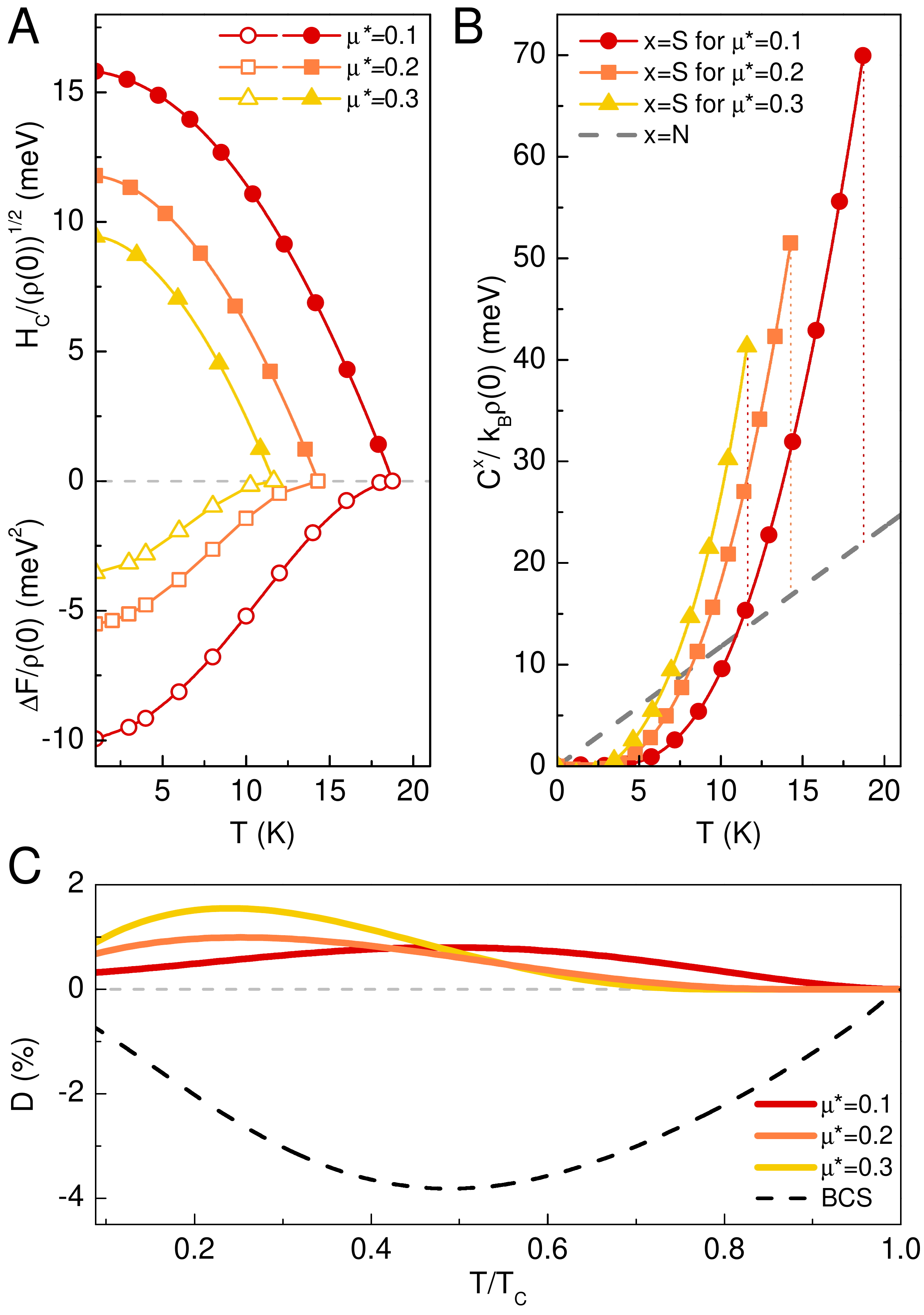}
\caption{(A) The free energy difference (lower panel) and the thermodynamic critical field (upper panel) as a function of temperature. (B) The specific heat of the superconducting and normal state as a function of temperature for selected values of Coulomb pseudopotential. (C) Critical field deviation as a function of the temperature.}
\label{fig3}
\end{figure}

The results obtained for silicene under the tension of 5$\%$ are presented in the lower panel of \fig{fig3} (A). From the physical point of view, the negative values of $\Delta F/\rho(0)$ inform us about the thermodynamic stability of the superconducting state in the temperature range from $T_0$ to $T_C$. It can be seen that the increase of the Coulomb pseudopotential substantially weakens the stability of the superconducting state in silicene.

The upper panel in \fig{fig3} (A) presents the influence of the temperature on the ratio $H_{C}/\sqrt{\rho\left(0\right)}$. It can be seen that the thermodynamic critical field decreases with the increasing temperature, taking the zero value for $T=T_{C}$. Let us notice that the maximum values of the considered function equal: $H_{C}\left(0\right)/\sqrt{\rho\left(0\right)}\in\left\langle 15.8, 9.4\right\rangle$ meV for $\mu^{\star}\in\left\langle0.1, 0.3\right\rangle$, where $H_{C}\left(0\right)\equiv H_{C}\left(T_{0}\right)$. 

\begin{figure*}[!t]
\includegraphics[width=2\columnwidth]{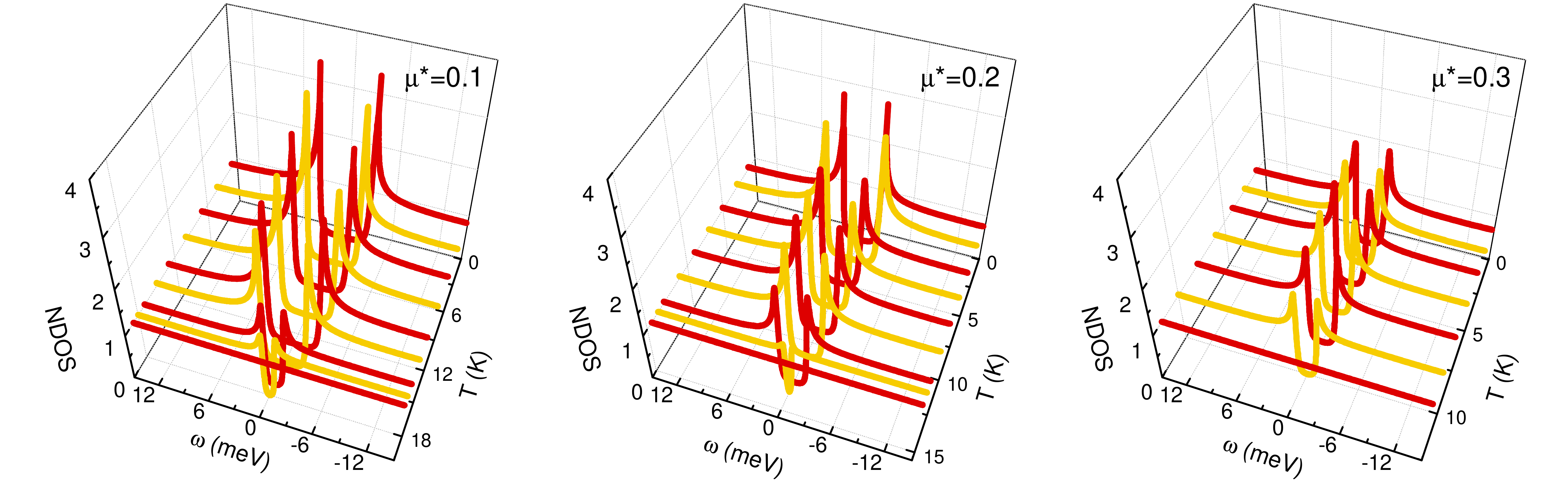}
\caption{The dependence of the total normalized density of states on the frequency for selected temperature and values of the Coulomb pseudopotential.}
\label{fig4}
\end{figure*}

In \fig{fig3} (B), the specific heat for the normal and superconducting state is presented. 
We can note that with the increase of the temperature the specific heat of the superconducting state grows strongly, reaching its maximum at the critical temperature.
The characteristic specific heat jump at the critical temperature is marked with the vertical line. 
We can observe that the value of the {\it jump}, just like the value of the thermodynamic critical field, decreases with the growth of the Coulomb pseudopotential. In particular this fact can be described by the following ratio: 
$\left[\Delta C\left(T_{C}\right)\right]_{\mu^{\star}=0.3}/ \left[\Delta C\left(T_{C}\right)\right]_{\mu^{\star}=0.1}=0.58$.

In \fig{fig3} (C) we supplement our results with the calculated thermodynamic critical field deviation as a function of the temperature for selected values of Coulomb pseudopotential. The positive values of $D$ function correspond to the strong electron-phonon coupling ($\lambda>1$) and $D$ is negative for the weak coupling limit ($\lambda<1$) \cite{Navarro}. Our results confirm that for the investigated material the electron-phonon coupling is strong. Results predicted by the Bardeen-Cooper-Schrieffer theory (BCS) \cite{BCS1}, \cite{BCS2} are presented with a dashed line \cite{Michor}.

In the next step, on the basis of the calculated thermodynamic functions, we determined the dimensionless ratios $R_{C}\equiv \Delta C\left(T_{C}\right)/C^{N}\left(T_{C}\right)$ and 
$R_{H}\equiv T_{C}C^{N}\left(T_{C}\right)/H^{2}_{C}\left(0\right)$. In the framework of the BCS model these parameters have the universal values: $\left[R_{C}\right]_{\rm BCS}=1.43$ and $\left[R_{H}\right]_{\rm BCS}=0.168$ \cite{BCS1}, \cite{BCS2}. Taking into account obtained results, we found that the values of $R_{C}$ and $R_{H}$ for silicene under the tension of 5$\%$ significantly diverge from the values predicted by the BCS model, in particular: $R_{C}\in\left\langle 2.19, 2.05\right\rangle$ and $R_{H}\in\left\langle 0.143, 0.155\right\rangle$.

The Eliashberg equations in the mixed representation are next solved for the identical range of the temperatures and the Coulomb pseudopotential as the Eliashberg equations on the imaginary axis.

On the basis of the solutions of the Eliashberg equations in the mixed representation the exact value of the order parameter can be obtained from the relation \cite{Cappelutti}: 
\begin{equation}
\label{r7}
\Delta\left(T\right)={\rm Re}\left[\Delta\left(\omega=\Delta\left(T\right),T\right)\right].
\end{equation}

The zeroth temperature superconducting energy gap at the Fermi level for the investigated case of silicene takes the following values: $2\Delta(0)\equiv2\Delta(T_{0})\in\left\langle6.68, 3.88\right\rangle$ meV for $\mu^{\star}\in\left\langle0.1, 0.3\right\rangle$.
The most interesting from the physical point of view is dimensionless ratio $R_{\Delta}\equiv2\Delta(0)/k_{B}T_{C}$.
In the framework of the BCS theory: $\left[R_{\Delta}\right]_{\rm BCS}=3.53$ \cite{BCS2}, 	
whereas in our case, even for a large values of $\mu^{\star}$, the ratio significantly exceeds the values predicted by the BCS model, in particular $R_{\Delta}\in\left\langle4.14, 3.87\right\rangle$.
This situation is caused by the fact that in the BCS model, the strong-coupling and retardation effects are not taken into account. Therefore this theory is valid only for values of the gap functions which are small compared to the Debye frequency. Strong-coupling and retardation effects are taken into consideration in Eliashberg theory and can be characterized by the ratio $k_{B}T_{C}/\omega_{ln}$. 
In the weak-coupling limit, one can assume: $\left[k_{B}T_{C}/\omega_{\rm ln}\right]_{\rm BCS}\rightarrow0$. For two-dimensional silicene layer under the tension of $5\%$ we obtained: $\left[k_{B}T_{C}/\omega_{\rm ln}\right]_{\mu^{\star}=0.1}\simeq 0.087$ and $\left[k_{B}T_{C}/\omega_{\rm ln}\right]_{\mu^{\star}=0.3}\simeq 0.054$.

Moreover, the order parameter function on the real axis allows to calculate the total normalized density of states \cite{Eliashberg}:
\begin{equation}
\label{r15}
{\rm NDOS}\left(\omega \right)=\frac{\rm DOS_{S}\left(\omega \right)}{\rm DOS_{N}\left(\omega \right)}={\rm Re}\left[\frac{\left|\omega -i\Gamma \right|}{\sqrt{\left(\omega -i\Gamma\right)^{2}}-\Delta^{2}\left(\omega\right)}\right],
\end{equation}
where the symbols $\rm DOS_{S}\left(\omega \right)$ and $\rm DOS_{N}\left(\omega \right)$ denote the density in the superconducting and normal state, respectively. The pair breaking parameter $\Gamma$ is equal to $0.15$ meV. 

The calculated total normalized density of states for different temperatures, and for selected values of Coulomb pseudopotential is presented in \fig{fig4}. These results allows us to trace the size of the superconducting gap with increasing values of the temperature and Coulomb pseudopotential. The symmetric maximas can be clearly observed for $T<T_C$. It is worth emphasizing, that the superconducting gap between the two symmetric maximas, is much larger than in the lithium-decorated graphene ($\rm LiC_6$) and is comparable with $\rm Li_2C_6$ \cite{Domin}. Above the critical temperature, the gap is no longer visible and silicene can reveals metallic behavior. 

In the last step, the wave function renormalization factor on the real axis ($Z\left(\omega\right)$) is determined.
In the framework of the Eliashberg formalism the real part of $Z\left(\omega\right)$ enables the determination of the electron effective mass ($m^{\star}_{e}$). In particular, the ratio of $m^{\star}_{e}$ to the electron band mass ($m_{e}$) is given by: 
$m^{\star}_{e}/m_{e}={\rm Re}\left[Z\left(0\right)\right]$. Let us notice that Re$\left[Z\left(0\right)\right]$, similarly as $Z_{m=1}$, takes the highest value for $T=T_{C}$. Thus, $\left[m^{\star}_{e}\right]_{{\rm max}}$ is equal to $2.11m_{e}$.
%

\section{Summary}

Using the Eliashberg approach, we have studied the behaviour of the superconducting critical temperature, energy gap, free energy difference between the superconducting and normal state, thermodynamic critical field and the specific heat in a strongly coupled electron-doped silicene under the tension of $5\%$.
The Coulomb pseudopotential effects on the thermodynamic properties have been extensively studied based on electron-phonon spectral function.
For $\mu^{\star}\in\left\langle0.1, 0.3\right\rangle$ the critical temperature and zeroth temperature superconducting energy gap at the Fermi level decrease from $18.7$ K to $11.6$ K and from $6.68$ meV to $3.88$ meV, respectively.  

Other thermodynamic parameters differ from the predictions of the BCS theory. In particular, for the dimensionless ratios of the calculated thermodynamic functions we reported the following results:
$R_{\Delta}\in\left\langle4.14, 3.87\right\rangle$, $R_{C}\in\left\langle 2.19, 2.05\right\rangle$ and $R_{H}\in\left\langle 0.143, 0.155\right\rangle$.
It is connected with fact that the Eliashberg formalism in contrast to BCS model does not omit the strong-coupling and retardation effects.

\begin{acknowledgments}

The authors are grateful to the Cz{\c{e}}stochowa University of Technology - MSK CzestMAN for granting access to the computing infrastructure built in the project No. POIG.02.03.00-00-028/08 "PLATON - Science Services Platform".

\end{acknowledgments}
\bibliographystyle{apsrev}
\bibliography{manuscript}
\end{document}